\documentclass[prd,twocolumn,showpacs,superscriptaddress,nofootinbib]{revtex4-1}
\usepackage[T1]{fontenc} 
\usepackage{amsmath}
\usepackage{amssymb}
\usepackage{amsfonts}
\usepackage{graphicx}
\usepackage{enumitem}
\usepackage{bm}
\usepackage{rotating}
\usepackage{hyperref}
\usepackage{xcolor}
\usepackage{array}
\usepackage{lmodern}
\usepackage[normalem]{ulem}
\usepackage{braket}
\usepackage{tensor}
\usepackage{mathrsfs}
\usepackage{float}

\usepackage[normalem]{ulem}

\begin{document}

\title{High-quality axion from chain seesaw}



\author{Pei-Hong Gu}

\email{phgu@seu.edu.cn}

\affiliation{School of Physics, Jiulonghu Campus, Southeast University, Nanjing 211189, China}

\begin{abstract}

We propose a chain seesaw mechanism to generate not only the masses of neutral neutrinos but also the masses of charged fermions, by supplementing the standard model $SU(3)_c^{}\times SU(2)_L^{}\times U(1)_Y^{}$ gauge symmetries with an additional $U(1)_{Y'}^{}$ gauge symmetry. The chains for the down-type and up-type quarks contain different numbers of links, because the corresponding mediator vector-like fermions and new Higgs scalars carry different $U(1)_{Y'}^{}$ charges. The Peccei-Quinn global symmetry is automatically embedded in this $U(1)_{Y'}^{} $ gauge symmetry which is sufficient to guarantee the high quality of axion.

\end{abstract}


\maketitle

\section{Introduction}


In the $SU(3)_c^{} \times SU(2)_L^{} \times U(1)_Y^{}$ standard model (SM) of particle physics, the charged fermions acquire their Dirac masses through the dimension-4 Yukawa interactions among the left-handed fermion doublets, the right-handed fermion singlets and the Higgs doublet. On the other hand, the right-handed neutrinos are absent from the SM, so the neutrinos remain massless within the SM. However, the discovery of neutrino oscillations firmly requires the existence of massive neutrinos. The most immediate way to generate the neutrino masses would be to introduce the right-handed neutrinos and then construct the dimension-4 Yukawa interactions, in analogy with the scheme for generating the charged fermion masses. Unfortunately, the ultralight neutrinos constrain their Yukawa couplings with the SM Higgs doublet to be extremely small, which is rather unnatural.

To naturally explain the tiny but nonzero neutrino masses, one may resort to the elegant seesaw mechanism \cite{minkowski1977,yanagida1979,grs1979,ms1980,mw1980,sv1980,cl1980,lsw1981,ms1981,flhj1989}, in which the SM is extended by certain lepton-number-violating terms involving heavy new particles. In these seesaw models, the neutrinos are of Majorana nature. However, the lepton number violation and hence the Majorana nature is only a theoretical assumption and has not been verified in any experiment so far. One has also applied the seesaw mechanism to Dirac neutrinos \cite{rw1983,rs1984,mp2002,gh2006}, in analogy with the Majorana neutrino case. In these Majorana or Dirac seesaw models, the cosmic baryon asymmetry, which poses another big challenge to the SM, can be produced through the so-called leptogenesis \cite{fy1986} or neutrinogenesis \cite{dlrw1999} mechanism. 

Frankly speaking, the neutrino mass generation and the charged fermion mass generation are vastly different in the framework of the SM and its seesaw extension. This mysterious difference inspires us to consider a universal seesaw scenario \cite{berezhiani1983,rajpoot1987,dw1987} where not only the neutrinos but also the charged fermions acquire their Dirac masses through the seesaw mechanism. The key step towards realizing the universal seesaw is to find a proper symmetry that forbids the dimension-4 Yukawa couplings with the SM Higgs doublet. The simplest symmetry of this kind may be a $U(1)_{Y'}^{}$ gauge symmetry, which also naturally motivates the introduction of three right-handed neutrinos \cite{cg2022}.

On the other hand, the QCD Lagrangian contains a term of CP-violating gluon density term, whose coefficient $\bar{\theta}$ is theoretically expected to be of order unity. However, the upper bound on the electric dipole moment of the neutron enforces the value of $\bar{\theta}$ to be extremely small, i.e. $|\bar{\theta}| < 10^{-10}_{}$. The unexplained huge gap between the theoretical expectation and the experimental result is referred to as the strong CP problem \cite{pq1977}. Currently, the most widely discussed solution to the strong CP problem is to introduce a pseudo Goldstone boson \cite{pq1977,weinberg1978,wilczek1978}, known as the axion, from the spontaneous breaking of a Peccei-Quinn (PQ) global symmetry $U(1)_{\textrm{PQ}}^{}$ \cite{pq1977} which is explicitly broken by the QCD anomaly \cite{adler1969,bj1969,ab1969}.

So far the axion has not been observed in any experiment. Such an invisible axion \cite{kim1979,svz1980,dfs1981,zhitnitsky1980} requires the $U(1)_{\textrm{PQ}}^{}$ global symmetry to be spontaneously broken at a very high scale, far above the weak scale. For a recent review, see \cite{dlgnv2020}. However, all global symmetries in nature are believed to be broken by non-perturbative gravity effects such as black holes and wormholes. In other words, the $U(1)_{\textrm{PQ}}^{}$ global symmetry would be explicitly broken by higher-dimensional operators suppressed by the Planck scale. Unless the coefficients of the relevant operators are extremely small, the axion would be significantly displaced from zero and would hence fail to remain the strong CP solution. The severe fine-tuning of these small coefficients, known as the axion quality problem \cite{mkjmr1992,hhkkww1992,sgmlmr1992,smbds1992}, has motivated a number of new mechanisms. For example, the $U(1)_{\textrm{PQ}}^{}$ global symmetry may emerge as an accidental symmetry of certain Abelian \cite{smbds1992,ycqjwwtty2023,ksbbdrnm2025} or non-Abelian \cite{ldiluzio2020,allstw2020} gauge symmetries.

In this paper we shall  propose a chain seesaw mechanism, based on a $U(1)_{Y'}^{}$ gauge symmetry, which generates not only the masses of the neutral neutrinos but also the masses of the charged fermions, meanwhile, while accommodating the neutrinogenesis processes. The chains for the down-type and up-type quarks contain different numbers of links, because the corresponding mediator vector-like fermions and new Higgs scalars carry different $U(1)_{Y'}^{}$ charges. Consequently, this $U(1)_{Y'}^{} $ gauge symmetry automatically includes a $U(1)_{\textrm{PQ}}^{}$ global symmetry. Moreover, the requirement of gauge invariance under this $U(1)_{Y'}^{}$ gauge symmetry is sufficient to ensure the axion quality.

\section{The model}


We conveniently classify the fermions into two groups: the ordinary fermions, which comprise the SM fermions and the right-handed neutrinos, as summarized in Table \ref{ofermions}, and the mediator fermions, which comprise the vector-like fermion singlets and the vector-like fermion doublets, as summarized in Table \ref{vectorfermions}. Clearly, the introduction of the right-handed neutrinos is necessary for the cancellation of the gauge anomalies, whereas the introduction of the vector-like fermions has no effect on the anomaly cancellation. The Higgs scalars, including the SM Higgs doublet and the other new Higgs singlets, are listed in Table \ref{hscalars}.

\begin{table}
\begin{center}
\begin{tabular}{|c|c|c|}  \hline &&\\[-2.0mm] $\begin{array}{c} Ordinary\\
[1mm]
fermions\end{array}$&~$SU(3)_c^{}\times SU(2)_L^{}\times U(1)_Y^{}$~&~$U(1)_{Y'}^{}$~\\
&&\\[-2.0mm]\hline&& \\[-1.5mm]
$q_L^{}$&$(3,2,+\frac{1}{6})$ & $-\frac{1}{4}$    \\
&&\\[-2.0mm]\hline&& \\[-1.5mm]$d_R^{}$ &$(3,1,-\frac{1}{3})$ &$-\frac{3}{4}$ \\
&&\\[-2.0mm]\hline&& \\[-1.5mm]$u_R^{}$ &$(3,1,+\frac{2}{3})$  & $+\frac{1}{4}$ \\
&&\\[-2.0mm]\hline&& \\[-1.5mm]$l_L^{}$ &$(1,2,-\frac{1}{2})$  & $+\frac{3}{4}$\\
&&\\[-2.0mm]\hline&& \\[-1.5mm]$e_R^{}$ &$(1,1,-1)$  & $+\frac{1}{4}$ \\
&&\\[-2.0mm]\hline&& \\[-1.5mm]$\nu_R^{}$ &$(1,1,0)$  & $+\frac{5}{4}$ \\
[1.5mm]
\hline
\end{tabular}
\vspace{0.25cm}
\caption{\label{ofermions} The ordinary fermions include the SM fermions and the right-handed neutrinos. The three family indices are not shown for simplicity. The introduction of three right-handed neutrinos is necessary for the cancellation of the gauge anomalies.}
\end{center}
\end{table}

\begin{table}
\begin{center}
\begin{tabular}{|c|c|c|}  \hline &&\\[-2.0mm] $\begin{array}{c}Vector\!-\!like\\
[1mm]
fermion\\
[1mm]
singlets\end{array}$&~$SU(3)_c^{}\times SU(2)_L^{}\times U(1)_Y^{}$~&~$U(1)_{Y'}^{}$~\\
&&\\[-2.0mm]\hline&& \\[-1.5mm]$D_{i}^{}~(i=1,...,m)$ &$(3,1,-\frac{1}{3})$  & $-\frac{1}{4}-\frac{i-1}{2m}$  \\
&&\\[-2.0mm]\hline&& \\[-1.5mm]$U_{i}^{}~(i=1,...,n)$ &$(3,1,+\frac{2}{3})$  & $-\frac{1}{4}+\frac{i-1}{2n}$ \\
&&\\[-2.0mm]\hline&& \\[-1.5mm]$E_{i}^{}~(i=1,...,p)$ &$(1,1,-1)$  & $+\frac{3}{4}-\frac{i-1}{2p}$  \\
&&\\[-2.0mm]\hline&& \\[-1.5mm]$N_{i}^{}~(i=1,...,q)$ &$(1,1,0)$  & $+\frac{3}{4}+\frac{i-1}{2q}$ \\
[1.5mm]
\hline\hline &&\\[-2.0mm] $\begin{array}{c} Vector\!-\!like\\
[1mm]
fermion\\
[1mm]
doublets\end{array}$&~$SU(3)_c^{}\times SU(2)_L^{}\times U(1)_Y^{}$~&~$U(1)_{Y'}^{}$~\\
&&\\[-2.0mm]\hline&& \\[-1.5mm]$\Psi_{i}^{}~(i=1,...,m)$ &$(3,2,+\frac{1}{6})$  & $-\frac{1}{4}-\frac{i}{2m}$  \\
&&\\[-2.0mm]\hline&& \\[-1.5mm]$\Omega_{i}^{}~(i=1,...,n)$ &$(3,2,+\frac{1}{6})$  & $-\frac{1}{4}+\frac{i}{2n}$ \\
&&\\[-2.0mm]\hline&& \\[-1.5mm]$\Sigma_{i}^{}~(i=1,...,p)$ &$(1,2,-\frac{1}{2})$  & $+\frac{3}{4}-\frac{i}{2p}$  \\
&&\\[-2.0mm]\hline&& \\[-1.5mm]$\Delta_{i}^{}~(i=1,...,q)$ &$(1,2,-\frac{1}{2})$  & $+\frac{3}{4}+\frac{i}{2q}$ \\
[1.5mm]
\hline
\end{tabular}
\vspace{0.25cm}
\caption{\label{vectorfermions} The mediator fermions include the vector-like fermion singlets and the vector-like fermion doublets. Specifically, $D_i^{}$ and $\Psi_i^{}$ are responsible for generating the down-type quark masses, $U_i^{}$ and $\Omega_i^{}$ for the up-type quark masses, $E_i^{}$ and $\Sigma_i^{}$ for the charged lepton masses, while $N_i^{}$ and $\Delta_i^{}$ are responsible for generating the neutrino masses. Here the indices denote the link indices rather than the family indices. For simplicity, the three family indices are not shown. }
\end{center}
\end{table}

\begin{table}
\begin{center}
\begin{tabular}{|c|c|c|}  \hline&&\\[-2.0mm] $\begin{array}{c} Higgs\\
 [1mm]
 scalars\end{array}$&~$SU(3)_c^{}\times SU(2)_L^{}\times U(1)_Y^{}$~&~$U(1)_{Y'}^{}$~\\
&&\\[-2.0mm]\hline&& \\[-1.5mm]$\phi$ &$(1,2,+\frac{1}{2})$  & $0$  \\
&&\\[-2.0mm]\hline&& \\[-1.5mm]$\xi_{d}^{}$ &$(1,1,0)$  & $+\frac{1}{2m}$ \\
&&\\[-2.0mm]\hline&& \\[-1.5mm]$\xi_{u}^{}$ &$(1,1,0)$  & $+\frac{1}{2n}$ \\
&&\\[-2.0mm]\hline&& \\[-1.5mm]$\xi_{e}^{}$ &$(1,1,0)$  & $+\frac{1}{2p}$ \\
&&\\[-2.0mm]\hline&& \\[-1.5mm]$\xi_{\nu}^{}$ &$(1,1,0)$  & $+\frac{1}{2q}$ \\
[1.5mm]
\hline
\end{tabular}
\vspace{0.25cm}
\caption{\label{hscalars} The Higgs scalars include the SM Higgs doublet and the other new Higgs singlets. The two Higgs singlets $\xi_{d}^{}$ and $\xi_{u}^{}$ are required to carry the different $U(1)_{Y'}^{}$ charges. }
\end{center}
\end{table}

Remarkably, the SM fermions and the right-handed neutrinos are assigned to carry the $U(1)_{Y'}^{}$ charges defined by $Y'=Y-(5/4)(B-L)$, where $Y$ denotes the SM hypercharges, $B$ the baryon number ($B=\frac{1}{3}$ for quarks and $B=0$ for leptons), while $L$ the lepton number ($L=0$ for quarks and $L=1$ for leptons). Under this convention, the SM Higgs doublet $\phi$, which carries neither baryon nor lepton number, would have a $U(1)_{Y'}^{}$ charge identical to its hypercharge. However, we deliberately exempt the SM Higgs doublet from this convention and instead assign no $U(1)_{Y'}^{}$ charge to it.

Unlike the SM Higgs doublet $\phi$, the new Higgs singlets $\xi_d^{}$, $\xi_u^{}$, $\xi_e^{}$ and $\xi_\nu^{}$ are charged under the $U(1)_{Y'}^{}$ gauge symmetry. As will be clarified below, the two Higgs singlets $\xi_{d}^{}$ and $\xi_{u}^{}$ must carry different $U(1)_{Y'}^{}$ charges in order to guarantee a high-quality axion. These two distinct Higgs singlets can be in one-to-one correspondence with the other two Higgs singlets $\xi_{e}^{}$ and $\xi_{\nu}^{}$. Alternatively, only one of the Higgs singlets $\xi_{d}^{}$ and $\xi_{u}^{}$ can be identified simultaneously with both Higgs singlets $\xi_{e}^{}$ and $\xi_{\nu}^{}$. Of course, these four Higgs singlets can also be different from one another. All possible choices of the Higgs singlets are listed as follows, 
\begin{eqnarray}
\left(\xi_d^{},\xi_u^{},\xi_e^{},\xi_\nu^{}\right) &= & \left\{\begin{array}{lcl} \!\! \!\! \!\left.\begin{array}{l}
\left(\xi_1^{},\xi_2^{},\xi_1^{},\xi_2^{}\right) \\
[2mm]
\left(\xi_1^{},\xi_2^{},\xi_2^{},\xi_1^{}\right)\\
[2mm]
\left(\xi_1^{},\xi_2^{},\xi_1^{},\xi_1^{}\right)\\
[2mm]
\left(\xi_1^{},\xi_2^{},\xi_2^{},\xi_2^{}\right)\end{array}\!\!  \right\} &\!\!\textrm{with}\!\!& \xi_1^{} \! \neq \! \xi_2^{}\,, \\
\\
\!\! \!\! \!\left.\begin{array}{l}\left(\xi_1^{},\xi_2^{},\xi_1^{},\xi_3^{}\right)\\
[2mm]
\left(\xi_1^{},\xi_2^{},\xi_2^{},\xi_3^{}\right)\\
[2mm]
\left(\xi_1^{},\xi_2^{},\xi_3^{},\xi_1^{}\right)\\
[2mm]
\left(\xi_1^{},\xi_2^{},\xi_3^{},\xi_2^{}\right)\end{array} \!\! \right\} &\!\!\textrm{with}\!\!& \xi_1^{}\! \neq \! \xi_2^{} \! \neq \!\xi_3^{}\,, \\
\\
\!\! \!\! \! \left.\begin{array}{l}\left(\xi_1^{},\xi_2^{},\xi_3^{},\xi_4^{}\right)\end{array}\!\! \right\}  &\!\!\textrm{with}\!\!& \xi_1^{}\! \neq \! \xi_2^{}\neq \xi_3^{} \! \neq \! \xi_4^{} \,.\end{array}\right. 
\end{eqnarray}

We then write down the allowed Yukawa and mass terms involving all fermions, i.e.
\begin{eqnarray}
\mathcal{L}_{Y+M}^{}&=& - y_{L}^{} \bar{f}_L^{}  F_{R1}^{} \varphi -\sum_{i=1}^{k-1} y_{i,i+1}^{}\bar{F}_{Li}^{}F_{Ri+1}^{} \xi  \nonumber\\
[2mm]
&& - y_R^{}\bar{F}_{Lk}^{} f_R^{} \xi- \sum_{i=1}^{k} M_{F_i}^{} \bar{F}_{Ri}^{} F_{Li}^{} +\textrm{H.c.} ~~\textrm{with}\nonumber\\
[2mm]
&&\left(f_L^{},f_R^{},\varphi,\xi,F_i^{}\right)=\left\{\begin{array}{l}\left(q_L^{},d_R^{},\phi,\xi_d^{},D_i^{}\right)\,, \\
[3mm]
\left(q_L^{},u_R^{},\tilde{\phi},\tilde{\xi}_u^{},U_i^{}\right)\,, \\
[3mm]
\left(l_L^{},e_R^{},\phi,\xi_e^{},E_i^{}\right)\,, \\
[3mm]
\left(l_L^{},\nu_R^{},\tilde{\phi},\tilde{\xi}_\nu^{},N_i^{}\right)\,,
\end{array}\right.
\end{eqnarray} 
for the vector-like fermion singlets, and  
\begin{eqnarray}
\mathcal{L}_{Y+M}^{}&=& - y_{L}^{} \bar{f}_L^{}  X_{R1}^{} \xi -\sum_{i=1}^{k-1} y_{i,i+1}^{}\bar{X}_{Li}^{}X_{Ri+1}^{}\xi  \nonumber\\
[2mm]
&&- y_R^{}\bar{X}_{Lk}^{} f_R^{} \varphi-\sum_{i=1}^{k} M_{X_i}^{} \bar{X}_{Ri}^{}X_{Li}^{}+\textrm{H.c.} ~~\textrm{with}\nonumber\\
[2mm]
&&\left(f_L^{},f_R^{},\varphi,\xi,X_i^{}\right)=\left\{\begin{array}{l}\left(q_L^{},d_R^{},\phi,\xi_d^{},\Psi_i^{}\right)\,, \\
[3mm]
\left(q_L^{},u_R^{},\tilde{\phi},\tilde{\xi}_u^{},\Omega_i^{}\right)\,, \\
[3mm]
\left(l_L^{},e_R^{},\phi,\xi_e^{},\Sigma_i^{}\right)\,, \\
[3mm]
\left(l_L^{},\nu_R^{},\tilde{\phi},\tilde{\xi}_\nu^{},\Delta_i^{}\right)\,,
\end{array}\right.
\end{eqnarray} 
for the vector-like fermion doublets. Here $\tilde{\phi}$, $\tilde{\xi}_{u}^{}$ and $\tilde{\xi}_{\nu}^{}$ are the charge-conjugated Higgs scalars defined by
\begin{eqnarray}
\tilde{\phi}=i\tau_2^{} \phi^\ast_{}\,,~~\tilde{\xi}_{u}^{}=\xi_{u}^\ast\,,~~\tilde{\xi}_{\nu}^{}=\xi_{\nu}^\ast\,.
\end{eqnarray}

\section{Chain seesaw}


Because of the present assignments of the $U(1)_{Y'}^{}$ charges, the ordinary SM fermions and right-handed neutrinos cannot directly couple to the SM Higgs doublet through the usual dimension-4 Yukawa interactions. However, with the participation of the new Higgs singlets, the down-type quarks, the up-type quarks, the charged leptons and the neutral neutrinos can interact with the SM Higgs doublet through the following effective operators, 
\begin{eqnarray}
\label{eff}
\mathcal{L}_{eff}^{}&=& -\sum_{i,j=1,2,3}^{}\left(\frac{c_{d}^{ij}}{\Lambda_d^{m}} \bar{q}_{Li}^{} \phi d_{Rj}^{} \xi_d^m +\frac{c_{u}^{ij}}{\Lambda_u^{n}} \bar{q}_{Li}^{} \tilde{\phi} u_{Rj}^{} \tilde{\xi}_u^n \right.\nonumber\\
[2mm]
&&\left.+\frac{c_{e}^{ij}}{\Lambda_e^{p}} \bar{l}_{Li}^{} \phi e_{Rj}^{} \xi_e^p +\frac{c_{\nu}^{ij}}{\Lambda_\nu^{q}} \bar{l}_{Li}^{} \tilde{\phi} \nu_{Rj}^{} \tilde{\xi}_\nu^q +\textrm{H.c.}\right)\,.
\end{eqnarray} 
After the Higgs singlets develop their VEVs, the ordinary fermions can obtain their dimension-4 Yukawa couplings to the SM Higgs doublet. When the Higgs doublet subsequently develops its VEV, the down-type quarks, the up-type quarks, the charged leptons and the neutral neutrinos can respectively acquire their Dirac masses as follows, 
\begin{eqnarray}
\label{fmass}
m_d^{} &=& \frac{c_d^{}\langle \phi\rangle \langle \xi_d^{}\rangle^{m}_{}}{\Lambda_d^{m}}\,,~~m_u^{} = \frac{c_u^{}\langle \phi\rangle \langle \xi_u^{}\rangle^{n}_{}}{\Lambda_u^{n}}\,,\nonumber\\
[2mm]
m_e^{} &=& \frac{c_e^{}\langle \phi\rangle \langle \xi_e^{}\rangle^{p}_{}}{\Lambda_e^{p}}\,,~~m_\nu^{} = \frac{c_\nu^{}\langle \phi\rangle\langle \xi_\nu^{}\rangle^{q}_{}}{\Lambda_\nu^{q}}\,.
\end{eqnarray}

\begin{figure*}
\centering
\includegraphics[scale=0.65]{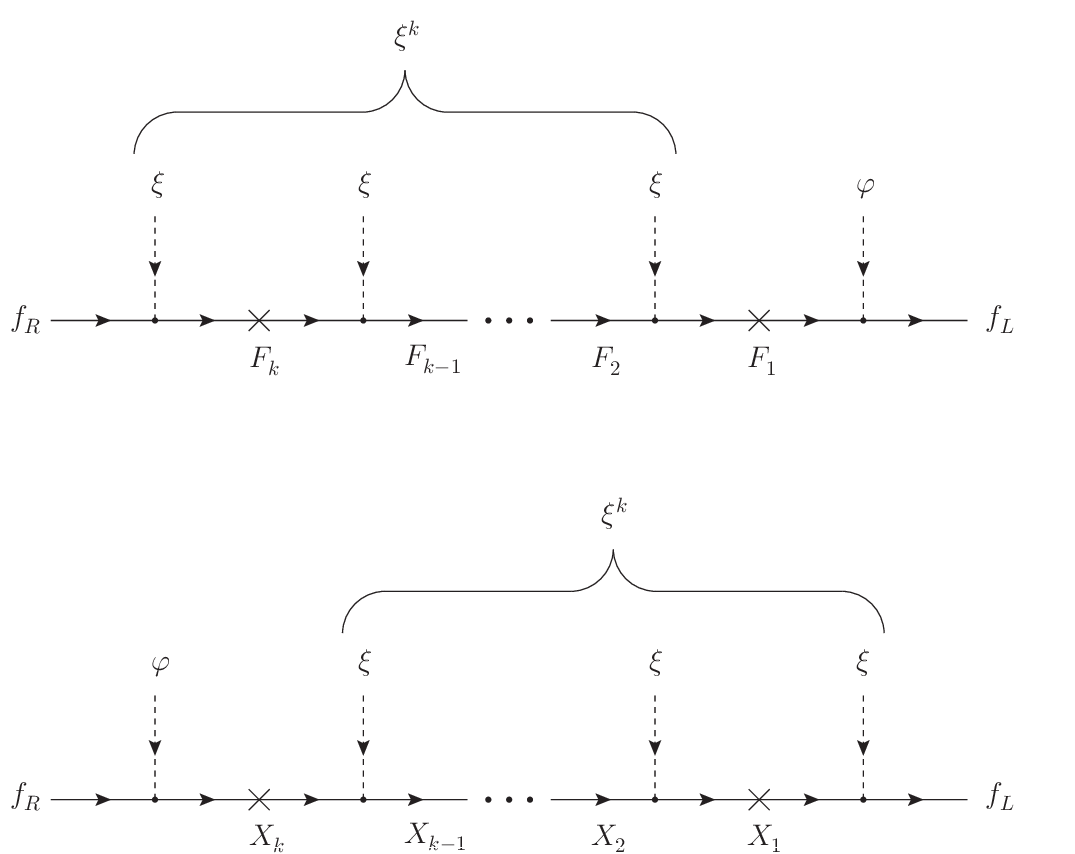} \caption{\label{chseesaw} The $U(1)_{Y'}^{}$ universal seesaw mediated by the fermion chains. Here $f_L^{}$ and $f_R^{}$ are the left-handed fermion doublets and the right-handed fermion singlets, respectively, while $\varphi$ and $\xi$ are the Higgs doublet and the Higgs singlet, respectively. Note that the Higgs scalars $\varphi$ and $\xi$ should be the Higgs doublet $\phi$ and the Higgs singlets $\xi_{d}^{}$ and $\xi_e^{}$ if the right-handed fermions $f_R^{}$ denote the down-type quarks $d_R^{}$ and the charged leptons $e_R^{}$. Otherwise, the Higgs scalars $\varphi$ and $\xi$ should be the charge-conjugated Higgs doublet $\tilde{\phi}=i\tau_2^{}\phi^\ast_{}$ and the charge-conjugated Higgs singlets $\tilde{\xi}_u^{}=\xi_u^\ast$ and $\tilde{\xi}_\nu^{}=\xi_\nu^\ast$ if the right-handed fermions $f_R^{}$ denote the up-type quarks $u_R^{}$ and the neutral neutrinos $\nu_R^{}$. As for the vector-like fermions $F_{i}^{}$ and $X_{i}^{}$, they are the iso-singlets and the iso-doublets, respectively. Specifically, we refer to $(f_L^{},f_R^{},F_{i }^{},X_{i}^{})$ as $(q_L^{},d_R^{},D_i^{},\Psi_i^{})$, $(q_L^{},u_R^{},U_i^{},\Omega_i^{})$, $(l_L^{},e_R^{},E_i^{},\Sigma_i^{})$ and $(l_L^{},\nu_R^{},N_i^{},\Delta_i^{})$, respectively.}
\end{figure*}

Actually, we can resort to the vector-like fermion singlets and/or doublets to realize these higher-dimensional operators in a renormalizable way. After ordering these vector-like fermions by their $U(1)_{Y'}^{}$ charges, the first and the last ones have the Yukawa couplings with the corresponding left-handed and right-handed ordinary fermions, while each pair of adjacent vector-like fermions can have a Yukawa coupling with the corresponding Higgs singlet. Consequently, these vector-like fermions form the chains that link the corresponding left-handed fermion doublets and right-handed fermion singlets at the two ends. Moreover, if the number of vector-like fermions in a chain is denoted by $k$, a total of $k+1$ Yukawa coupling coefficients enter this chain. Eventually, the corresponding dimensionless coefficients in the effective operators (\ref{eff}) are given by the products of these $k+1$ Yukawa coupling coefficients, while the $k$-th power of the cutoff is the product of the masses of these vector-like fermions on this chain. This means that the hierarchy among the masses of the ordinary fermions, including the neutrinos, can be naturally accommodated either by the $k+1$ Yukawa couplings or by the $k$ vector-like fermion masses.

The above statement can also be understood with the help of the Feynman diagrams shown in Fig. \ref{chseesaw}. Since the route from the right-handed fermion singlets to the left-handed fermion doublets is mediated by the chains composed of several vector-like fermions, this mechanism for generating the Yukawa couplings of the ordinary fermions to the SM Higgs doublet may be figuratively named the chain seesaw.

We further clarify that the chain seesaw can accommodate a neutrinogenesis mechanism to explain the baryon asymmetry in the present universe. For example, on the fermion chain that generates the neutrino masses, the $j$-th link of vector-like fermions is assumed to be heavier than the $(j\pm 1)$-th links, so that its out-of-equilibrium and CP-violating decays can generate a lepton asymmetry stored in the $(j-1)$-th link of vector-like fermions and an equal but opposite lepton asymmetry stored in the $(j+1)$-th link of vector-like fermions. As the $(j-1)$-th link of vector-like fermions eventually decays into the left-handed lepton doublets, its lepton asymmetry can be converted into the lepton asymmetry stored in the left-handed leptons. On the other hand, as the $(j+1)$-th link of vector-like fermions eventually decays into the right-handed neutrino singlets, its lepton asymmetry can be converted into the lepton asymmetry stored in the right-handed neutrinos. Because their effective Yukawa couplings to the SM Higgs doublet are too weak, the left-handed lepton doublets and the right-handed neutrino singlets can only go into equilibrium at the very low temperatures, where the electroweak sphaleron processes \cite{krs1985} have already stopped converting the lepton asymmetry stored in the left-handed leptons into the baryon asymmetry stored in the quarks. Therefore, only the lepton asymmetry produced in the left-handed leptons can be partially converted into a baryon asymmetry through the electroweak sphaleron processes. For simplicity, we shall not delve into more details regarding the generation of the lepton and baryon asymmetries here.

\section{High-quality axion}


It is easy to see that we can obtain a QCD axion from the effective operators (\ref{eff}) if the Higgs singlets $\xi_d^{}$ and $\xi_u^{}$ are two independent Higgs scalars. Specifically, this axion is given by  
\begin{eqnarray}
a= \frac{m \langle\xi_u^{}\rangle \textrm{Im}(\xi_d^{}) - n \langle\xi_d^{}\rangle \textrm{Im}(\xi_u^{}) }{\sqrt{n^2_{}\langle\xi_d^{}\rangle^2_{}+ m^2_{}\langle\xi_u^{}\rangle^2_{}}} \,,
\end{eqnarray}
and its decay constant is determined by 
\begin{eqnarray}
f_a^{} =\frac{mn \langle\xi_d^{}\rangle \langle\xi_u^{}\rangle}{\sqrt{n^2_{}\langle\xi_d^{}\rangle^2_{}+ m^2_{}\langle\xi_u^{}\rangle^2_{}}} \,.
\end{eqnarray}
On the other hand, it is believed that all global symmetries of nature should be broken by non-perturbative gravity effects such as black holes and wormholes. This implies that certain higher-dimensional operators suppressed by the Planck scale would explicitly break the $U(1)_{\textrm{PQ}}^{}$ global symmetry in the Lagrangian. These effects would displace the axion significantly away from zero and would spoil the strong CP solution, unless the dimension of the relevant operators turns out to be high enough. 

The simplest way to distinguish one Higgs singlet $\xi_d^{}$ from the other Higgs singlet $\xi_u^{}$ is to assume that these two Higgs singlets can carry the same $U(1)_{Y'}^{}$ charge but two different $U(1)_{\textrm{PQ}}^{}$ charges, just like the two Higgs doublets in the original PQ model \cite{pq1977,weinberg1978,wilczek1978}. Therefore, we should additionally impose a proper discrete symmetry to guarantee a high enough dimension for the lowest-order gravity-induced PQ symmetry breaking term, i.e.
\begin{eqnarray}
\mathcal{O}=\frac{1}{\left(k!\right)^2_{}}\frac{ \left(\xi_d^{} \tilde{\xi}_u^{} \right)^k_{}}{M_{\textrm{Pl}}^{2k-4}} +\textrm{H.c.}\,.
\end{eqnarray}
Here and hereafter $M_{\textrm{Pl}}^{} =2.4\times 10^{18}_{}\,\textrm{GeV}$ is the reduced Planck mass.

A more attractive scheme is that the Higgs scalars $\xi_d^{}$ and $\xi_u^{}$ carry two different $U(1)_{Y'}^{}$ charges and hence automatically carry two different $U(1)_{\textrm{PQ}}^{}$ charges. Without assuming any additional discrete symmetry, the $U(1)_{Y'}^{}$ gauge symmetry itself can ensure a sufficiently high dimension for the lowest-order gravity-induced PQ symmetry breaking term, i.e.
\begin{eqnarray}
\mathcal{O}=\frac{1}{m! n! }\frac{\xi_d^m \tilde{\xi}_u^n }{M_{\textrm{Pl}}^{m+n-4}} +\textrm{H.c.}\,.
\end{eqnarray}
This means we can naturally realize a high-quality axion \cite{ycqjwwtty2023,ksbbdrnm2025}.

\section{Conclusion}


In this paper, we have demonstrated that both the neutral neutrinos and the charged fermions can obtain their Dirac masses through a universal chain seesaw mechanism based on a $U(1)_{Y'}^{}$ gauge symmetry. At the same time, the cosmic baryon asymmetry can be produced by a neutrinogenesis mechanism within this context. Moreover, the number of links on the chain for the down-type quarks can remain different from the number of links on the chain for the up-type quarks, as the corresponding mediator vector-like fermions and new Higgs scalars carry different $U(1)_{Y'}^{}$ charges. Due to the requirement of gauge invariance, this $U(1)_{Y'}^{} $ gauge symmetry, which automatically contains the $U(1)_{\textrm{PQ}}^{}$ global symmetry, can be sufficient to guarantee the high quality of axion.

\textbf{Acknowledgement}: This work was supported in part by the National Natural Science Foundation of China under Grant No. 12175038.

\end{document}